\documentclass[]{spie}  

\usepackage{amsmath,amsfonts,amssymb}
\usepackage{graphicx}
\usepackage{multicol}
\usepackage{tabu}
\usepackage[colorlinks=true, allcolors=blue]{hyperref}
\usepackage{float}
\usepackage{gensymb}

\title{Detector Characterization of a Near-Infrared Discrete Avalanche Photodiode 5x5 Array for Astrophysical Observations}

\author[a]{Siyang Li}
\author[b]{J\'er\^ome Maire}
\author[b,c]{Maren Cosens}
\author[b,c]{Shelley A. Wright}
\affil[a]{Department of Physics, University of California Berkeley, USA}
\affil[b]{Center for Astrophysics $\&$ Space Sciences, University of California San Diego, USA}
\affil[c]{Department of Physics, University of California San Diego, USA}

\authorinfo{Further author information: (Send correspondence to S.L.)\\S.L.: E-mail: seanli@berkeley.edu \\ }

\begin{document} 
\maketitle

\begin{abstract}
We present detector characterization of a state-of-the-art near-infrared (950nm - 1650 nm) Discrete Avalanche Photodiode detector (NIRDAPD) 5x5 array. We designed an experimental setup to characterize the NIRDAPD dark count rate, photon detection efficiency (PDE), and non-linearity. The NIRDAPD array was illuminated using a 1050 nm light-emitting diode (LED) as well as 980 nm, 1310 nm, and 1550 nm laser diodes. We find a dark count rate of 3.3x10\textsuperscript{6} cps, saturation at 1.2x10\textsuperscript{8} photons per second, a photon detection efficiency of 14.8$\%$ at 1050 nm, and pulse detection at 1 GHz. We characterized this NIRDAPD array for a future astrophysical program that will search for technosignatures and other fast ($>$ 1 Ghz) astrophysical transients as part of the Pulsed All-sky Near-infrared Optical Search for Extraterrestrial Intelligence (PANOSETI) project. The PANOSETI program will consist of an all-sky optical (350 - 800 nm) observatory capable of observing the entire northern hemisphere instantaneously and a wide-field NIR (950 - 1650 nm) component capable of drift scanning the entire sky in 230 clear nights. PANOSETI aims to be the first wide-field fast-time response near-infrared transient search.

\end{abstract}


\keywords{SETI, technosignatures, near-infrared, avalanche photodiode, astrobiology, instrumentation, telescopes, astrophysical transients, observational astronomy
}

\section{INTRODUCTION}
\label{sec:intro}  

One of the most effective means of interstellar communication is the laser\cite{Schwartz61}, which with today's technology can produce pulses up to petawatts in power and picoseconds in duration \cite{Hatchett}. If used for interstellar communication on Earth, these lasers would outshine our own sun by at least 4 orders of magnitude and be bright enough to be easily distinguished from natural astrophysical sources by civilizations with high resolution meter class telescopes from thousands of light years away. \cite{Horowitz04}

Most Searches for Extraterrestrial Intelligence (SETI) have focused on the radio and visible spectra \cite{Drake61, Siemion2010, Werthimer2001, Howard2000, Guillermo93}. However, the transmission of near-infrared signals over large distances in the Galactic plane can be advantageous over the transmission of both radio and optical signals due to lower extinction factors through ionized interstellar medium \cite{Wright14} and negligible pulse width distortion from scattering \cite{Horowitz04}, suggesting that near-infrared lasers could also be used by intelligent extraterrestrial civilizations for interstellar communication. Previous near-infrared astrophysical surveys and SETI programs on Earth have been limited by the timing resolutions of modern photodetectors. The introduction of the InGaAs/InP Single Photon Avalanche Detector (SPAD) has opened up the possibility of probing the near-infrared Universe in the nanosecond regime to search for and characterize fast ($>$1 GHz) transient events. These novel detectors operate between 950 nm and 1650 nm and are p-n junction reverse-biased semiconductor detectors that utilize avalanche multiplication \cite{Linga10}. The Near-Infrared Discrete Avalanche Photodetector (NIRDAPD) from Amplification Technologies is a SPAD that uses Internal Discrete Amplification (IDA) technology and surpasses previous InGaAs/InP SPADs with reported faster responses ($>$  1 GHz), higher gains (1x10\textsuperscript{5}), higher photon detection efficiencies (15$\%$ at 1550 nm) and lower noise ($\sim$ 10\textsuperscript{6} counts per second) \cite{Linga10}.

The Pulsed All-sky Near-infrared Optical Search for Extraterrestrial Intelligence (PANOSETI) project aims to probe the largely unexplored near-infrared nanosecond regime using state-of-the-art NIRDAPD technology by searching for technosignatures and other astrophysical transients. The  PANOSETI program will consist of an all-sky optical (350 - 800 nm) observatory capable of instantaneously observing the entire northern hemisphere and a wide-field near-infrared (950 - 1650 nm) component capable of drift scanning the entire sky in 230 clear nights. PANOSETI aims to be the first wide-field fast-time response near-infrared technosignature search.

Two single pixel NIRDAPDs from Amplification Technologies were commissioned on the 1 meter Nickel telescope at Lick Observatory for the Near Infrared and Optical Search for Extraterrestrial (NIROSETI) project, the predecessor to PANOSETI, in March 2015. \cite{ Wright14}. NIROSETI has the sensitivity to detect laser emissions from up to 50 parsecs away and has since observed 2,000 celestial objects\cite{Maire16}. 

We characterized the dark count rate, linearity, saturation, photon detection efficiency, and pulse detection of a 5x5 NIRDAPD array (1550 series) from Amplification Technologies to explore the feasibility of replacing the current single pixel NIRDAPD detectors at Lick Observatory with NIRDAPD arrays to increase our search area and sensitivity. We also evaluate the feasibility of using NIRDAPD arrays for other observational programs searching for fast near-infrared signals of astrophysical origin.

\section{Experimental Setup}

\begin{figure} [ht]
\begin{center}
\begin{multicols}{2}
\includegraphics[height=6cm]{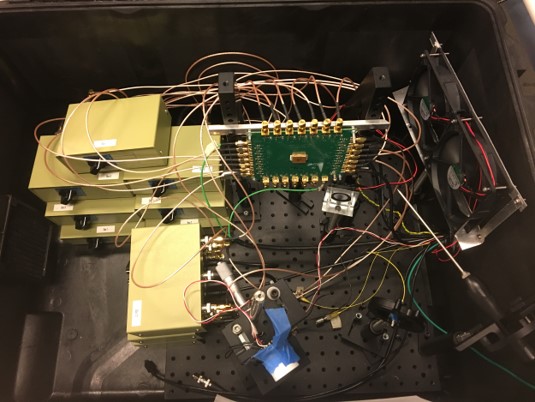}\par 
\includegraphics[height=6cm]{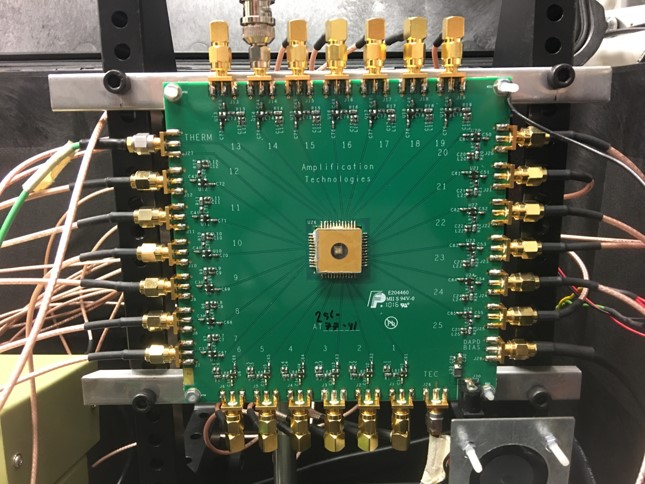}\par 
\end{multicols}
\end{center}
\caption{LEFT: Interior of the dark box setup used to characterize the NIRDAPD array. BNC switches were used to connect the analog outputs of the NIRDAPD to an oscilloscope and can be seen in the left half of the dark box. The NIRDAPD array and cooling system can be seen in the right half of the dark box. RIGHT: The NIRDAPD 5x5 array characterized in this study.}
\label{fig:darkbox}
\end{figure}

\begin{figure} [H]
\begin{center}
\includegraphics[height=8cm]{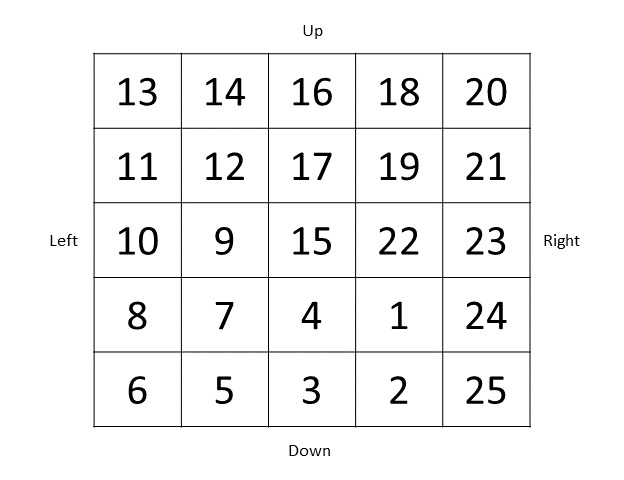}\par 
\end{center}
\caption{Pixel locations and orientation. This diagram will be used to reference the location of pixels throughout this paper.}
\label{fig:pixellayout}
\end{figure}

The experimental setup can be seen in Figure \ref{fig:darkbox}. The NIRDAPD array was housed inside a light tight dark box  containing a 25.4 mm diameter sealed cable feed-through. Darkness at 950 nm and 1050 nm was verified throughout the box using a power meter (ThorLabs PM100D) with a sensitivity of 0.1 nW and an accuracy of $\pm$0.5$\%$. The NIRDAPD array was hermetically sealed inside a TO-8 can, mounted on a breakout board supplied by Amplification Technologies, and cooled using a thermoelectric cooler (ThorLabs TTC001) and thermistor. Detector parameters can be seen in Table \ref{tab:parameters}. The pixel layout and orientation can be seen in Figure  \ref{fig:pixellayout} and will be used to reference the location of pixels throughout this paper.

A single count was defined as a pulse with an amplitude greater than half the amplitude produced by a single photoelectron (threshold = 0.5 mV). Counts per second were obtained using a 4 channel 2.5 GHz oscilloscope (Agilent MSO9254A) with MATLAB installed. A two layered system consisting of ten four-way BNC switch boxes was constructed to change pixels while minimizing disturbances to the NIRDAPD array between measurements. Measurements were taken three pixels at a time, leaving the fourth oscilloscope channel connected directly to an arbitrary pixel (Pixel 13) to monitor for any changes in counts per second over time. A cooling system consisting of fans and radiators was constructed to remove excess heat inside the dark box produced by the breakout board. The NIRDAPD array and amplifier were allowed to settle for at least 30 minutes before taking dark count measurements and 90 minutes before taking measurements with a light source.

\begin{table}[H]
\begin{center}
\begin{tabu} to 0.8\textwidth { | X[c] | X[c] | }

 \hline
 \textbf{Parameter} & \textbf{Value} \\
 \hline
 Operating Bias  & -61.1V \\
  \hline
 Amplifier Bias  & +12V  \\
  \hline
 Operating Temperature  & 250 K  \\
  \hline
 Total Detector Area & 500 x 500 microns  \\
  \hline
 Pixel Dimension  & 100 x 100 microns  \\
  \hline
 Pixel Pitch  & 81 $\%$  \\
 \hline

\end{tabu}

\end{center}
\caption{Operating parameters and characteristics of the NIRDAPD 5x5 array.}
\label{tab:parameters}
\label{tab:template}
\end{table}

\section{Results}

\subsection{Dark Count Rate}

We found an average dark count rate of 3.3x10$\textsuperscript{6}$ counts per second per pixel across the array over five trials, which corresponds to a noise-equivalent power of 0.62 picowatts per pixel. The standard deviation was 3.7x10$\textsuperscript{5}$ cps. We observed a hot spot in the lower right hand corner of the array centered around pixel 2. Pixel 2 had a dark count rate 22$\%$ higher than the average, and pixel 20 had a dark count rate 22$\%$ lower than the average.

\begin{figure} [H]
\begin{center}
\begin{multicols}{2}
\includegraphics[height=6.5cm]{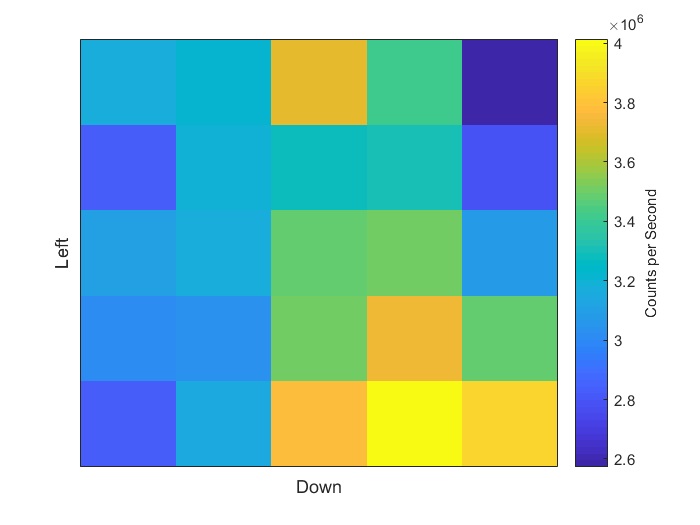}\par
\includegraphics[height=7cm]{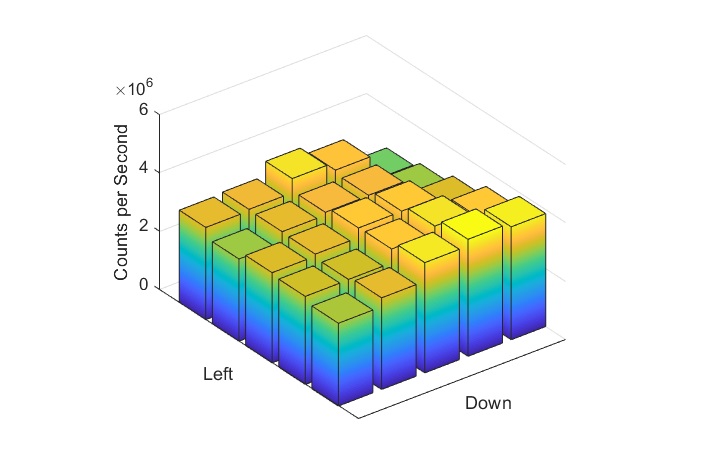}\par
\end{multicols}
\end{center}
\caption{Both images show the same set of data. We found an average dark count rate of 3.3x10$\textsuperscript{6}$ counts per second over five trials across the entire array. LEFT: A 2D pixel map showing the average dark count rate distribution across the NIRDAPD array. RIGHT: A 3D pixel map showing the average dark count rate distribution across the NIRDAPD array. }
\label{fig:darkcount}
\end{figure}

\subsection{Linearity and Saturation}

\begin{figure} [H]
\begin{center}
\includegraphics[height=8cm]{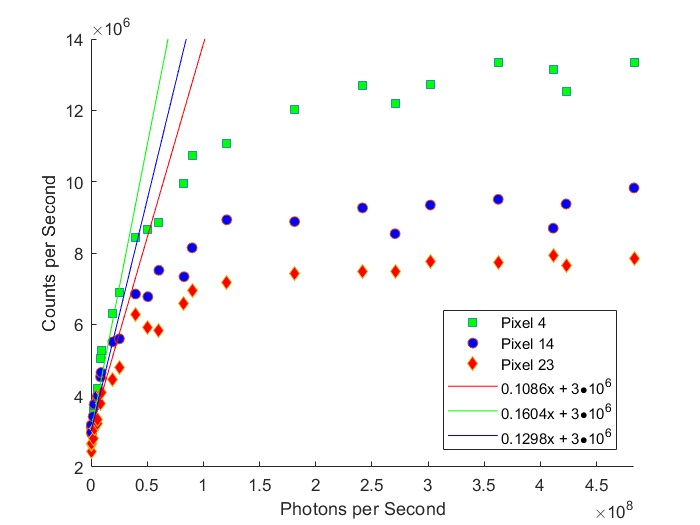}\par 
\end{center}
\caption{Counts per second as a function of incident photons per second for three random pixels (4, 14, and 23). The best fit line to the linear response region of each pixel is shown as a solid line.}
\label{fig:linearity}
\end{figure}

We observed the linearity for three random pixels (4, 14, and 23) at 1050 nm (Fig. \ref{fig:linearity}). Photon flux was varied by placing neutral density filters with optical densities of 0.1, 0.3, 0.6, 1.0, and 2.0 in front of the light source. The outputs of the three NIRDAPD pixels were observed to increase linearly with incident photon flux until deviating from the line of best fit by 10$\% $ at 3x10\textsuperscript{7} photons per second. Pixels 14 and 23 began saturating at approximately 1.2x10\textsuperscript{8} photons per second and pixel 4 began saturating at approximately 2.5x10\textsuperscript{8} photons per second. 

\subsection{Photon Detection Efficiency}

We obtained the PDE for each pixel at 980 nm, 1050 nm, 1310 nm, and 1550 nm at an incident power of 5.7 nW. The PDE was calculated by dividing the number of photons received by the detector $N_{detector}$ by the number of photons incident on the detector $N_{incident}$ and using the equation

\begin{equation}
    PDE (\%)= \frac{N_{detector}}{N_{incident}} =  \frac{N_{counts}-N_{dark}}{P_{incident}} \frac{A_{power meter}}{A_{NIRDAPD}} \frac{hc}{\lambda}
\end{equation}

where $N_{counts}$ is the total counts per second outputted by the NIRDAPD, $N_{dark}$ is the dark count rate, $P_{incident}$ is the incident power on the NIRDAPD measured by the power meter, ${A_{power meter}}$ is the area of the photosensitive surface of the power meter, ${A_{NIRDAPD}}$ is the photosensitive area of each pixel in NIRDAPD array, h is Planck's constant, c is the speed of light, and $\lambda$ is the wavelength of light incident on the NIRDAPD.

\begin{figure*} [ht]
\begin{center}
\begin{multicols}{2}
\includegraphics[height=6cm]{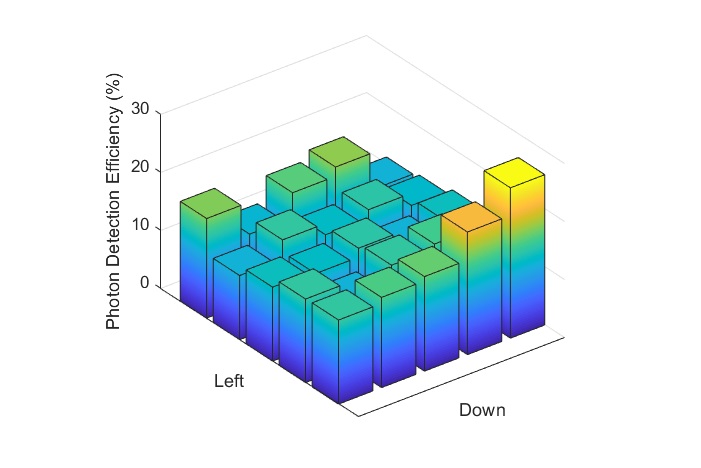}\par 
\includegraphics[height=6cm]{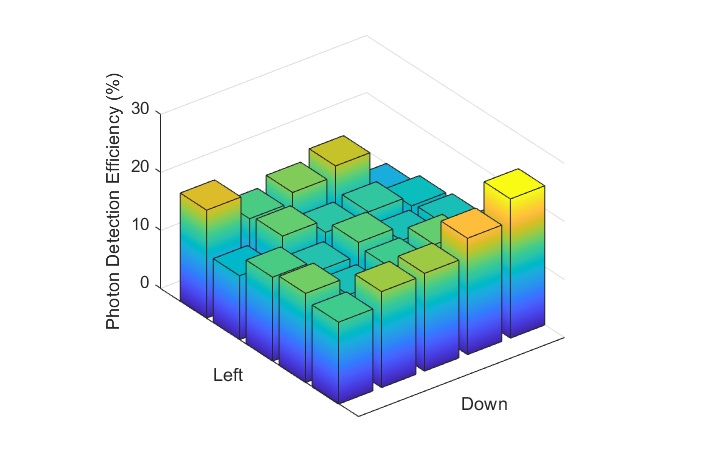}\par 
\end{multicols}
\begin{multicols}{2}
\includegraphics[height=6cm]{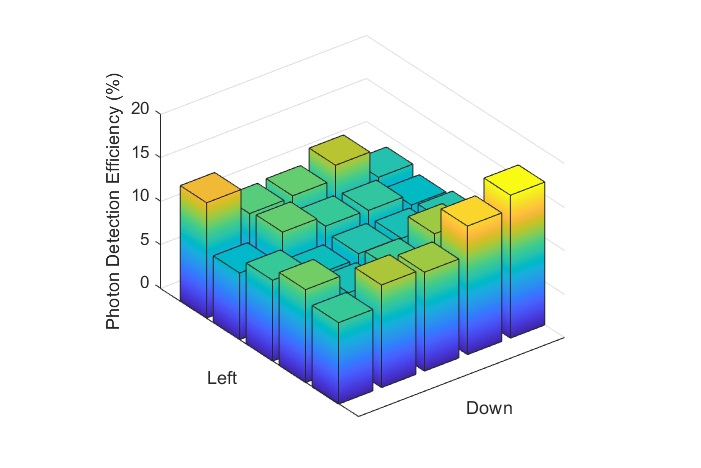}\par 
\includegraphics[height=6cm]{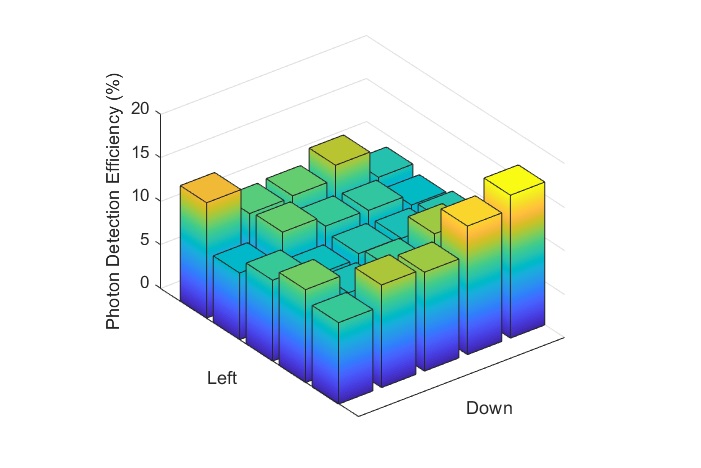}\par 
\end{multicols}
\end{center}
\caption{3D pixel maps showing the average PDE over 3 trials at various wavelengths. TOP LEFT: 980 nm, TOP RIGHT: 1050 nm, BOTTOM LEFT: 1310 nm, BOTTOM RIGHT: 1550 nm.}
\label{fig:PDE1}
\end{figure*}

\begin{figure} [H]
\begin{center}
\begin{tabular}{c} 
\includegraphics[height=8cm]{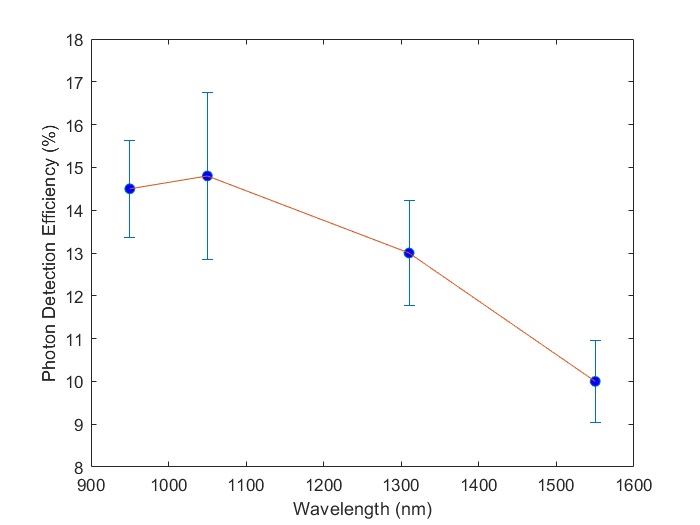}
\end{tabular}
\end{center}
\caption[example] 
{ \label{fig:example} 
Average PDE over all 25 pixels measured at 950 nm, 1050 nm, 1050 nm, 1310 nm and 1550.}
\label{fig:PDE2}
\end{figure}

The temperature of the detector was kept constant to $\pm$0.1 \degree C throughout each measurement of 25 pixels. Pixel maps of the average PDE over three trials at 980 nm, 1050 nm, 1310 nm, and 1550 nm can be seen in Figure \ref{fig:PDE1}, and the average PDE across all pixels as a function of wavelength can be seen in Figure \ref{fig:PDE2}. 

We observed pixel 25 to have a 68$\%$ higher PDE than average for each wavelength. This trend was consistent throughout two days of measurements and not caused by external noise sources.

\subsection{Pulse Detection}

\begin{figure} [H]
\begin{center}
\includegraphics[height=7.5cm]{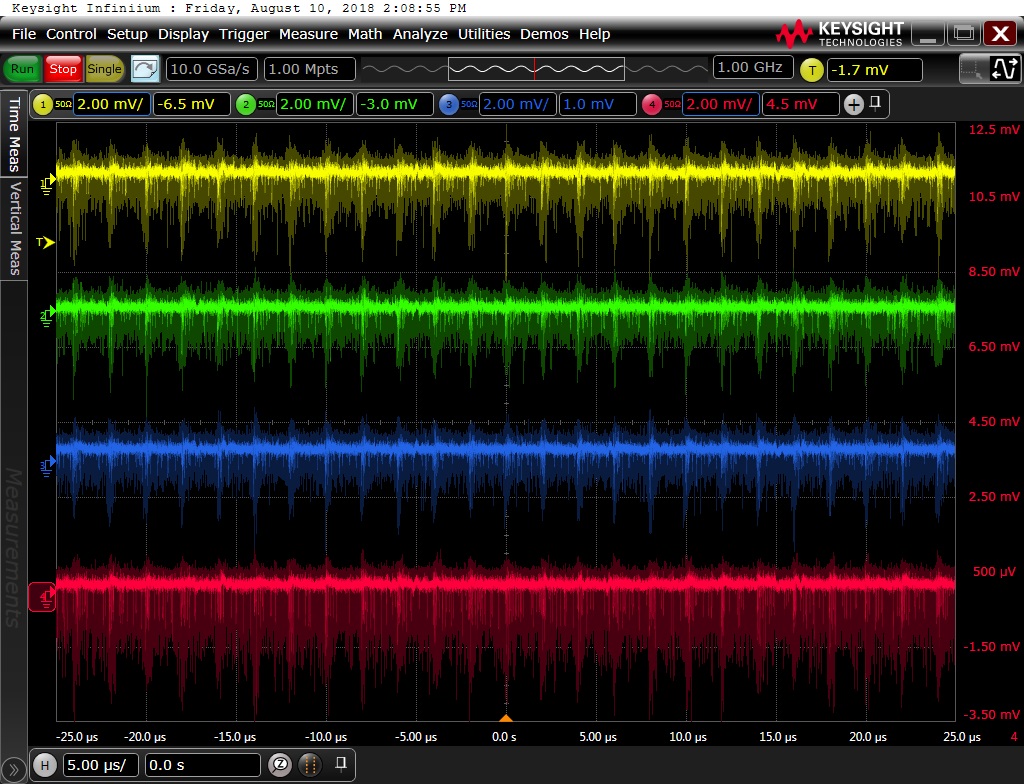}\par 
\end{center}
\caption{Response of four random pixels to a simulated technosignature produced by a 1550 nm laser diode pulsed at 500 kHz.}
\label{fig:pulse}
\end{figure}

We simulated a technosignature by modulating a 1550 nm laser diode with a waveform generator (Keysight 33210A). A buffer amplifier (Mini Circuits ZX60-6013E-S+) was powered at +12V and placed between the laser diode and the waveform generator to match impedances. As we increased the intensity of the laser, we observed areas of increased pulse amplitudes spaced at regular intervals corresponding to the frequency of the modulated laser diode. We were able to discriminate 1 ns width pulses and confirmed the 1 GHz bandwidth of the NIRDAPD array. The responses of four random pixels to the laser diode pulsed at 500 kHz can be seen in Figure \ref{fig:pulse}.

\section{Conclusion}

We characterized the dark count rate, linearity, saturation, and photon detection efficiency as well as confirmed the 1 GHz bandwidth of a NIRDAPD 5x5 array for the PANOSETI observatory. We created dark count rate pixel maps over five trials and found an average dark count rate of 3.7x10\textsuperscript{5} cps at 250K, which is consistent with the data provided by the manufacturer. We discovered a hot spot located in the lower right hand corner of the array centered at Pixel 2 which could be caused by a thermal gradient induced by the heat dissipation of nearby amplifiers. Identifying the cause of this anomaly will be the subject of future testing. 

We also created PDE pixel maps at 980 nm, 1050 nm, 1310 nm, and 1550 nm and made the first PDE measurements at  980 nm, 1050 nm, and 1310 nm for this NIRDAPD array. We found a peak PDE approaching 15$\%$ at 1050 nm. We characterized the non linearity and saturation for three random pixels by measuring the output of the NIRDAPD array as a function of incident photon flux. Pixels 4, 14, and 23 began deviating from linearity by 10$\%$ at 3x10\textsuperscript{7} counts per second. Pixels 14 and 23 saturated at 1.2x10\textsuperscript{8} photons per second while Pixel 4 saturated at 2.5x10\textsuperscript{8} photons per second. We also confirmed pulse detection capabilities of the NIRDAPD array by pulsing a 1550 nm laser diode and simulating a technosignature.

As we anticipate a sky background of 2.5x10\textsuperscript{6} counts per second at Lick Observatory and observed a peak PDE of approximately 15$\%$, we expect each pixel of the NIRDAPD to see a maximum of approximately 3.8x10\textsuperscript{5} photons per second and operate linearly and below the saturation limit during observations.

From the characteristics obtained in this study, we find that this NIRDAPD 5x5 array meets the requirements needed for the PANOSETI program and is a suitable near-infrared photodetector to use for astrophysical observations at Lick Observatory. The results of this study are not limited to the PANOSETI observatory and suggest this NIRDAPD array would also be a promising candidate detector for other astrophysical surveys searching for fast near-infrared transients. 

\acknowledgments 

The PANOSETI research and instrumentation program is made possible by the enthusiastic support and interest by Franklin Antonio. We thank the Bloomfield Family Foundation for supporting SETI research at UC San Diego in the CASS Optical and Infrared Laboratory.”

\bibliography{report} 

\begin{thebibliography}{10}

\bibitem{Schwartz61}
{Schwartz}, R.~N. and {Townes}, C.~H., ``{Interstellar and Interplanetary
  Communication by Optical Masers},'' {\em Nature}~{\bf 190},  205--208 (Apr.
  1961).

\bibitem{Hatchett}
{Hatchett}, S.~P., {Brown}, C.~G., {Cowan}, T.~E., {Henry}, E.~A., {Johnson},
  J.~S., {Key}, M.~H., {Koch}, J.~A., {Langdon}, A.~B., {Lasinski}, B.~F.,
  {Lee}, R.~W., {Mackinnon}, A.~J., {Pennington}, D.~M., {Perry}, M.~D.,
  {Phillips}, T.~W., {Roth}, M., {Sangster}, T.~C., {Singh}, M.~S., {Snavely},
  R.~A., {Stoyer}, M.~A., {Wilks}, S.~C., and {Yasuike}, K., ``{Electron,
  photon, and ion beams from the relativistic interaction of Petawatt laser
  pulses with solid targets},'' {\em Physics of Plasmas}~{\bf 7},  2076--2082
  (May 2000).

\bibitem{Horowitz04}
Howard, A.~W., Horowitz, P., Wilkinson, D.~T., Coldwell, C.~M., Groth, E.~J.,
  Jarosik, N., Latham, D.~W., Stefanik, R.~P., Jr.., A. J.~W., Wolff, J., and
  Zajac, J.~M., ``{Search for Nanosecond Optical Pulses from Nearby Solar-Type
  Stars},'' {\em The Astrophysical Journal}~{\bf 613},  1270--1284 (Oct. 2004).

\bibitem{Drake61}
{Drake}, F.~D., ``{Project Ozma},'' {\em Physics Today}~{\bf 14},  40--46
  (1961).

\bibitem{Siemion2010}
{Siemion}, A., {Von Korff}, J., {McMahon}, P., {Korpela}, E., {Werthimer}, D.,
  {Anderson}, D., {Bower}, G., {Cobb}, J., {Foster}, G., {Lebofsky}, M., {van
  Leeuwen}, J., and {Wagner}, M., ``{New SETI sky surveys for radio pulses},''
  {\em Acta Astronautica}~{\bf 67},  1342--1349 (Dec. 2010).

\bibitem{Werthimer2001}
Werthimer, D., Anderson, D., Bowyer, C.~S., Cobb, J., Heien, E., Korpela,
  E.~J., Lampton, M.~L., Lebofsky, M., Marcy, G.~W., McGarry, M., and Treffers,
  D., ``Berkeley radio and optical seti programs: {SETI}@home, {SERENDIP}, and
  {SEVENDIP},'' in [{\em The Search for Extraterrestrial Intelligence (SETI) in
  the Optical Spectrum II}{\nolinebreak\hspace{0.1em}]},  Kingsley, S.~A. and
  Bhathal, R., eds., {\em Proc. SPIE} {\bf 4273},  104--109 (Aug. 2001).

\bibitem{Howard2000}
{Howard}, A., {Horowitz}, P., {Coldwell}, C., {Klein}, S., {Sung}, A., {Wolff},
  J., {Caruso}, J., {Latham}, D., {Papaliolios}, C., {Stefanik}, R., and
  {Zajac}, J., ``{Optical SETI at Harvard-Smithsonian},'' in [{\em Bioastronomy
  99}{\nolinebreak\hspace{0.1em}]},  {Lemarchand}, G. and {Meech}, K., eds.,
  {\em Astronomical Society of the Pacific Conference Series} {\bf 213} (2000).

\bibitem{Guillermo93}
{Lemarchand}, G.~A., {Beskin}, G.~M., {Colomb}, F.~R., and {Mendez}, M.,
  ``{Radio and optical SETI from the southern hemisphere},'' in [{\em The
  Search for Extraterrestrial Intelligence (SETI) in the Optical
  Spectrum}{\nolinebreak\hspace{0.1em}]},  {Kingsley}, S.~A., ed., {\em Proc.
  SPIE} {\bf 1867},  138--154 (Aug. 1993).

\bibitem{Wright14}
{Wright}, S.~A., {Werthimer}, D., {Treffers}, R.~R., {Maire}, J., {Marcy},
  G.~W., {Stone}, R.~P.~S., {Drake}, F., {Meyer}, E., {Dorval}, P., and
  {Siemion}, A., ``{A near-infrared SETI experiment: instrument overview},'' in
  [{\em Ground-based and Airborne Instrumentation for Astronomy
  V}{\nolinebreak\hspace{0.1em}]},  {\em Proc. SPIE} {\bf 9147},  91470J (July
  2014).

\bibitem{Linga10}
{Linga}, K., {Yevtukhov}, Y., and {Liang}, B., ``{High-gain and low-excess
  noise near-infrared single-photon avalanche detector arrays},'' in [{\em
  Advanced Photon Counting Techniques IV}{\nolinebreak\hspace{0.1em}]},  {\em
  Proc. SPIE} {\bf 7681},  76810X (Apr. 2010).

\bibitem{Maire16}
{Maire}, J., {Wright}, S.~A., {Dorval}, P., {Drake}, F.~D., {Duenas}, A.,
  {Isaacson}, H., {Marcy}, G.~W., {Siemion}, A., {Stone}, R.~P.~S., {Tallis},
  M., {Treffers}, R.~R., and {Werthimer}, D., ``{A near-infrared SETI
  experiment: commissioning, data analysis, and performance results},'' in
  [{\em Ground-based and Airborne Instrumentation for Astronomy
  VI}{\nolinebreak\hspace{0.1em}]},  {\em Proc. SPIE} {\bf 9908},  990810 (Aug.
  2016).

\end{thebibliography}
\bibliographystyle{spiebib} 

\end{document}